\documentclass[12pt,preprint]{aastex}
\shorttitle{Wind velocity at 200$\,$mb}
\shortauthors{Carrasco et al.}
\begin{document}
\slugcomment{Accepted in PASP January 25, 2005}
\title{\bf High altitude wind velocity at  Sierra Negra and San Pedro M\'artir }

\author{Esperanza Carrasco}

\affil{Instituto Nacional de  Astrof\'{\i}sica, \'Optica y Electr\'onica,\\ 
Luis Enrique Erro \# 1, Tonantzintla, Puebla 72840, Mexico}
\email{bec@inaoep.mx}

\author{Remy Avila\altaffilmark{1}} 
\affil{Instituto de Astronom\'\i a, UNAM, M\'exico D.F.,Mexico \altaffiltext{1}{On leave from Centro 
de Radioastronom\'{\i}a y 
Astrof\'{\i}sica, UNAM, Morelia}}
\email{r.avila@astrosmo.unam.mx} 

\and

\author{Alberto Carrami\~nana}
\affil{Instituto Nacional de  Astrof\'{\i}sica, \'Optica y Electr\'onica,\\ 
Luis Enrique Erro \# 1, Tonantzintla, Puebla 72840, Mexico}
\email{alberto@inaoep.mx}

{}

\begin{abstract} {It has been  proposed that the  global circulation of the atmosphere winds at 200~mb 
can be used as a criteria to establish  the suitability  of a site for the development of  adaptive
optics techniques such as slow wavefront corrugation correction. By using the  NOAA NCEP/NCAR Reanalysis 
data base we analyze the monthly  average wind velocity at  200$\,$mb  for a 16 year period, for two sites
in Mexico: Sierra Negra and San Pedro M\'artir. We compare the results with  those obtained for 
Mauna Kea, Paranal and La Silla,  with Maidanak in Uzbekistan,
and with Gamsberg in Namibia. We show 
that for all the sites under study there is a yearly  wind speed modulation and we model that
modulation. Our results show that Sierra Negra and San Pedro M\'artir
are comparable with 
the
best observatory sites as Mauna Kea  and are amongst themost advantageous sites
to apply 
adaptive 
optics techniques.}

\end{abstract}

\section{INTRODUCTION}
   
\subsection{The relevance of the high altitute winds}

     One of the similarities of the next generation telescopes, also called Extremely Large Telescopes,
  is their reliance on adaptive optics. Therefore the  selection of the sites for the new generation 
  telescopes demands reliable studies of the turbulence vertical profiles, represented by the refractive 
  index structure  constant $C_{N}^{2}$, and the velocity of the turbulance layers ${\bf V}(h)$.  
  The average  velocity of the turbulance  V$_{0}$, defined as function of $C_{N}^{2}$ and
   ${\bf V}(h)$, is a  key  parameter  because most  quantities  relevant to adaptive optics 
  are directly  related to  V$_{0}$ (Roddier,  Gilli  \& Lund  1982).  Thus, if it is 
  possible to obtain  V$_{0}$ by a independent and reliable mean, the rest of the adaptive optics 
  quantities can be estimated 
from V$_{0}$.

      Sarazin \& Tokovinin  (2002) have shown that for two sites in the north of Chile, Paranal, the 
  site of the Very Large Telescope  and for Cerro Pach\'on, the site of Gemini South, the average
  velocity of the turbulence V$_{0}$ is clearly related to the wind velocity at 200$\,$mb pressure
  level. Assuming a standard  atmosphere, 200$\,$mb correspond to about  12 km  above sea  level. For
  their analysis they include data from 35 balloon flights.  They  compared the V$_{0}$ obtained
  from the  in situ measuremens with those obtained at 200$\,$mb from the NOAA (National Oceanic
  and Atmospheric Administration) Global Gridded  Upper Air database (GGUAS).  The result is the linear
  relation V$_{0}$ = $Max(v_{ground}, 0.4 v_{200 mb})$, with a rms=2.6  m$\,s^{-1}$, where $v_{ground}$ is the
  wind  velocity at ground level. The authors conclude that with moderate wind at ground level and 
  an equal seeing quality, the existing and potential  astronomical sites can be ranked in terms of
  suitability for adaptive optics simply by looking at the wind speed at 200$\,$mb.

    Hence, a first evaluation of  a site  convinience for developing  adaptive optics techniques
  can be the statistics of the wind speed at 200$\,$mb.  Moreover, the use of meteorological data bases 
  for astronomical characterization  offers  the possibility of  carrying  out studies, under the 
  same conditions, of different sites. Here, we present the statistics of the  200$\,$mb wind 
  velocity  monthly  variation for the Mexican sites Sierra Negra and  San Pedro M\'artir. We 
  compare the results  with those obtained for  Mauna Kea, Hawaii,  Paranal and La Silla in Chile, 
  Gamsberg in Namibia and  Maidanak in Uzbekistan.  A detailed study about the
  Canary Islands observatories can be found in Chueca et al. (2004). It must be emphasized 
  that we are not presenting a full characterization of the sites. Nevertheless we remark  
  the quality of our sites regarding  the high latitute wind speed as a very important 
  parameter for  adaptive optics.

\subsection{Sierra Negra and San Pedro M\'artir sites}

       Sierra Negra is  the site of the Large Millimeter Telescope (LMT/GTM), 
a 50~m antenna optimized  for 1-3~mm observations, now under construction.  It is located
at 18$^{\circ}$~59$^{'}06^{"}$~N latitude, 97$^{\circ}$~18$^{'}53^{"}$~W  longitude and 
an altitude of 4580m above sea level, in the state of Puebla, Mexico. Sierra Negra is about 100~km West 
from the coast of Veracruz in the Gulf of Mexico and 300~km from the Pacific Coast, within the
National Park Pico de Orizaba.  The LMT is  a bi-national project between Mexico
and the United States, led by the Instituto Nacional de Astrof\'{\i}sica, \'Optica y Electr\'onica
(INAOE) in Mexico and the University of Massachusetts, at Amherst, in the USA. I

     With the development of the Sierra Negra site,  the INAOE started to characterize  its quality for 
observations in other wavelengths. Because of its altitude and location, the site is intrinsically
very  dry, the opacity  at 240~GHz goes  down to $\la 0.02$, (Meza~et~al. 2003) therefore the 
conditions are likely 
to be favorable for near/mid infrared observations. A median seeing of 0.73 arcsec  
was obtained  by using a Differential Image Motion Monitor (DIMM) during 127 nights.  The data spans
between  February 2000 and November 2003 (Carrasco~et~al 2003a,  Avil\'es 2004). The meteorological 
conditions  are quite mild for such a high altitude, snowfall is generally light throughout the year. The 
global median temperature is  1.2$^\circ$C, the diurnal cycle is tipically only 2$^\circ$C  and the average 
temperature varies seasonally about 3$^\circ$C. The  global median wind velocity is  4  m$\,s^{-1}$, the  first 
and third quartile are 2.2  m$\,s^{-1}$ and 5.8  m$\,s^{-1}$, respectively with a diurnal wind speed  median tipically 
of 3.6  m$\,s^{-1}$. The average wind speed varies seasonally less than 2  m$\,s^{-1}$ (Carrasco et al. 2003b).

     San Pedro M\'artir (SPM) is the site of the Observatorio Astron\'omico Nacional,  located
 at 31$^{\circ}$02$^{'}$46$^{"}$~N latitute, 115$^{\circ}$~29$^{'}13^{"}$~W longitude and
 a altitude of 2830m above sea level, in  Baja California state, Mexico. It is at about 60~km from the
 Pacific and 53~km from the Gulf of Cort\'es. The site, administrated by the Universidad 
 Nacional Aut\'onoma de M\'exico (UNAM), is within a National Park.  It has three optical 
 telescopes: 2~m, 1.5~m  and a 0.84~m diameter. A study carried out between 1982 and 2002, shows that 
 the fraction of  photometric nights was  64\%  while the fraction of  spectroscopic nights  was 
 81\% (Tapia 2003). Futhermore, Tapia remarks that the fractional number  of totally clear, photometric 
 nights  increased considerably, from 54.2\%  prior to 1996 to 74.6\%  in the next seven years. The median 
 visual extinction is $k_{y}=0.13$ mag/airmass (Schuster, Parrao \& Guichard 2002, Parrao \& Schuster 2003). Atmospheric turbulence profiles
 above the observatory site have been obtained during several campaigns (Avila et al. 2003, 
 Avila et al. 2004a and Avila et al. 2004b). A median  seeing of 0.60 arcsec  has been obtained with 
 a  DIMM during  123 observing nights  between 2000 and  2003 (Michel et al. 2003). The wind speed median 
 is 3.9  m$\,s^{-1}$ during the day and 5.3  m$\,s^{-1}$ at night time (Michel, Hiriart \& Chapela 2003).  
 In fact, the site characterization carried out over the years
has proven that 
 SPM is indeed a very good site for optical/NIR  observations  and  one of the best sites 
 (Cruz-G\'onzalez, Avila \& Tapia 2003).

 \section{THE DATA}

        The  NOAA in the USA runs several climatological projects.  Within the NOAA, the  National
   Climatic 
Data Center (NCDC) is in charge of  managing the resource of global 
climatological
  in-situ and remotely sensed data and  information. Weather data
 from the atmosphere  are 
  obtained from instrument packages such as radiosondes
and  rawinsondes  carried   by weather 
  balloons that transmit the data back to a receiving station on the ground. The upper air 
  data consists of temperature, relative humidity, atmospheric pressure and wind.

       In a similar way the NOAA Climate Diagnostics Center (CDC) goal is to identify
  the nature and causes for climate variations on time scales ranging from a month to
  centuries. The CDC NCEP/NCAR   (National Center for Environmental Prediction/National Center
  for Atmospheric Research) Global Reanalysis Project is using the state-of-the-art 
  analysis and forecast system to perform data assimilation using past data from 1948 
  to the present. The  NCEP's role is to use a current and fixed (Jan. 1995) 
  version of a data assimilation and operational forecast  model.  The  
  task of NCAR is to collect and organize many of the land and marine surface data archives, 
  to provide these to NCEP along with many observed satelite and aircraft observations, receive 
  and store the  output archives.  The CDC Derived NCEP Reanalysis products  include over 
   80 different variables and  several
different coordinates systems at 0Z, 6Z, 12Z, and 
   18Z forecasted values.
  In particular,  the  Derived NCEP Pressure Level product provides, 
   the monthly
 wind speed on a 2.5 degree global grid at 17 pressure levels, with about 200 km
   horizontal resolution (Kistler et al. 2001). We are
using the monthly wind velocity  
   at 200$\,$mb on the basis of four records per day between 1980 and 1995.

     For  our analysis we used the NCEP/NCAR Reanalysis data base because it provides a more 
  accurate determination
of the monthly average wind speed than  the GGUAS, used by 
  Sarazin \& Tokovinin (2002), the 
fluctuations are  more than a factor of two 
  larger  for the 
GGUAS than for the NCEP data (Carrasco \& Sarazin 2003).

 \section{Results}
   
      The  coordinates of the sites under study  are shown
in table \ref{sites}. Costa Rica
 (San Jos\'{e}) is included as a tropical place where there is not jet stream. Two sets of  
 coordinates
are given  for each site. The first ones are the  geographic coordinates used 
 as
input
to the data base,  the next two columns  correspond to the grid points closest to 
 the geographical 
coordinates. The elevation in meters above sea level is included in the last column.

      In  Tables \ref{ncep_1} and  \ref{ncep_2}  the   wind velocities obtained  from the NCEP/NCAR   
 reanalysis data base are shown. The first column indicates the month number. For each site
 two values are reported. First, the wind speed  obtained by averaging the monthly values
 over the 16 years period. Second, the instantaneous rms fluctuations around the average. 
For the whole period, the annual average and  the quadratic average of the monthly
rms values are  included.  The latter represents a typical fluctuation of the wind speed.

\begin{figure}
\plotone{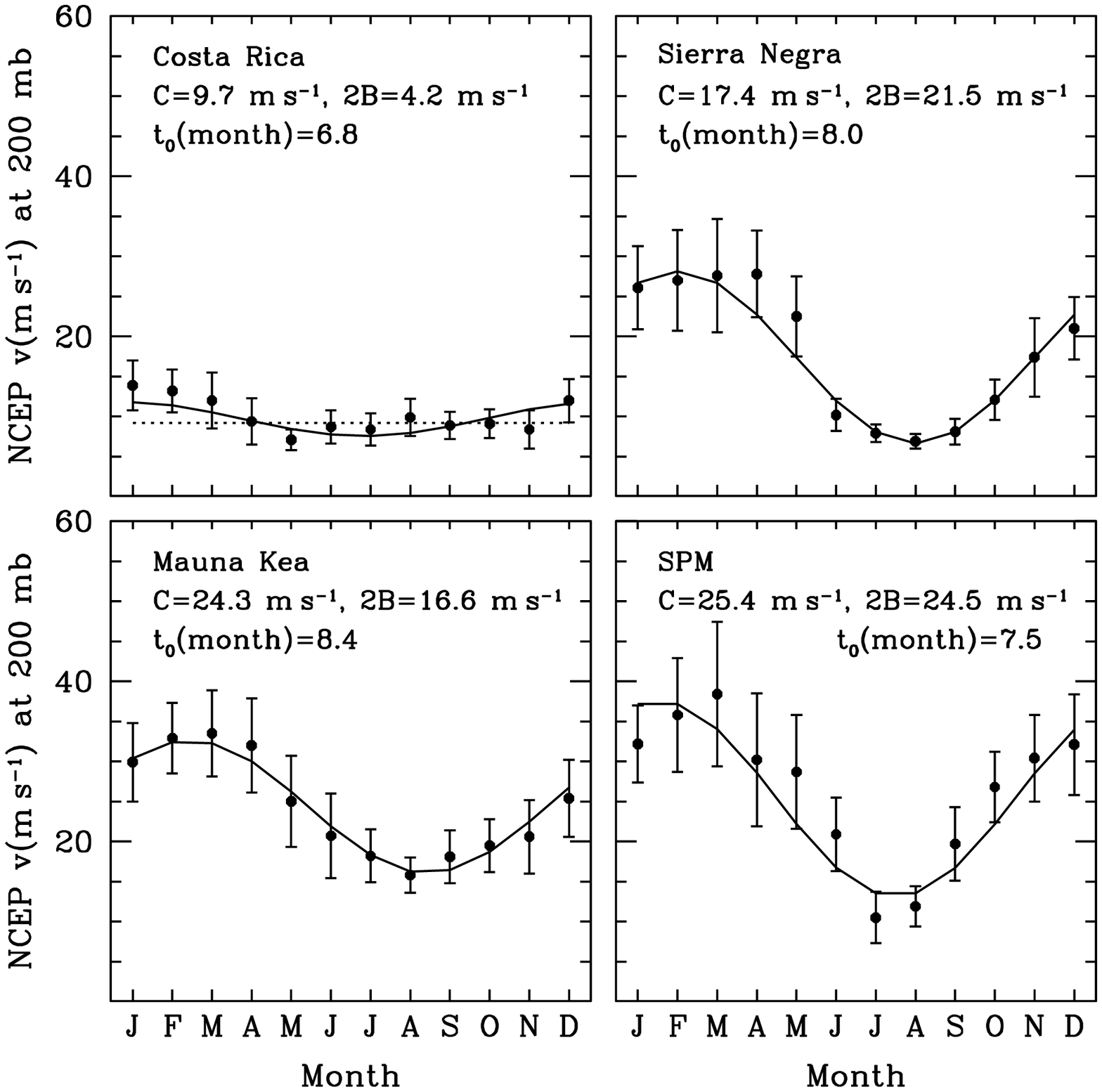}
\caption{Monthly wind speed for the sites considered. Averages and rms 
deviations are indicated by the points and corresponding error bars. The solid 
lines indicate the best fit of a yearly modulated wind speed, 
$v(t) = C_{s} - B\cos\{2\pi (t-t_{0})/1~{\rm year}\}.$ Dotted lines indicate a 
constant wind speed fit, $v(t)=C_{c}$, shown only for the data which do not reject
such hypothesis. \label{ncep_fit1}}
\end{figure}

\begin{figure}
\plotone{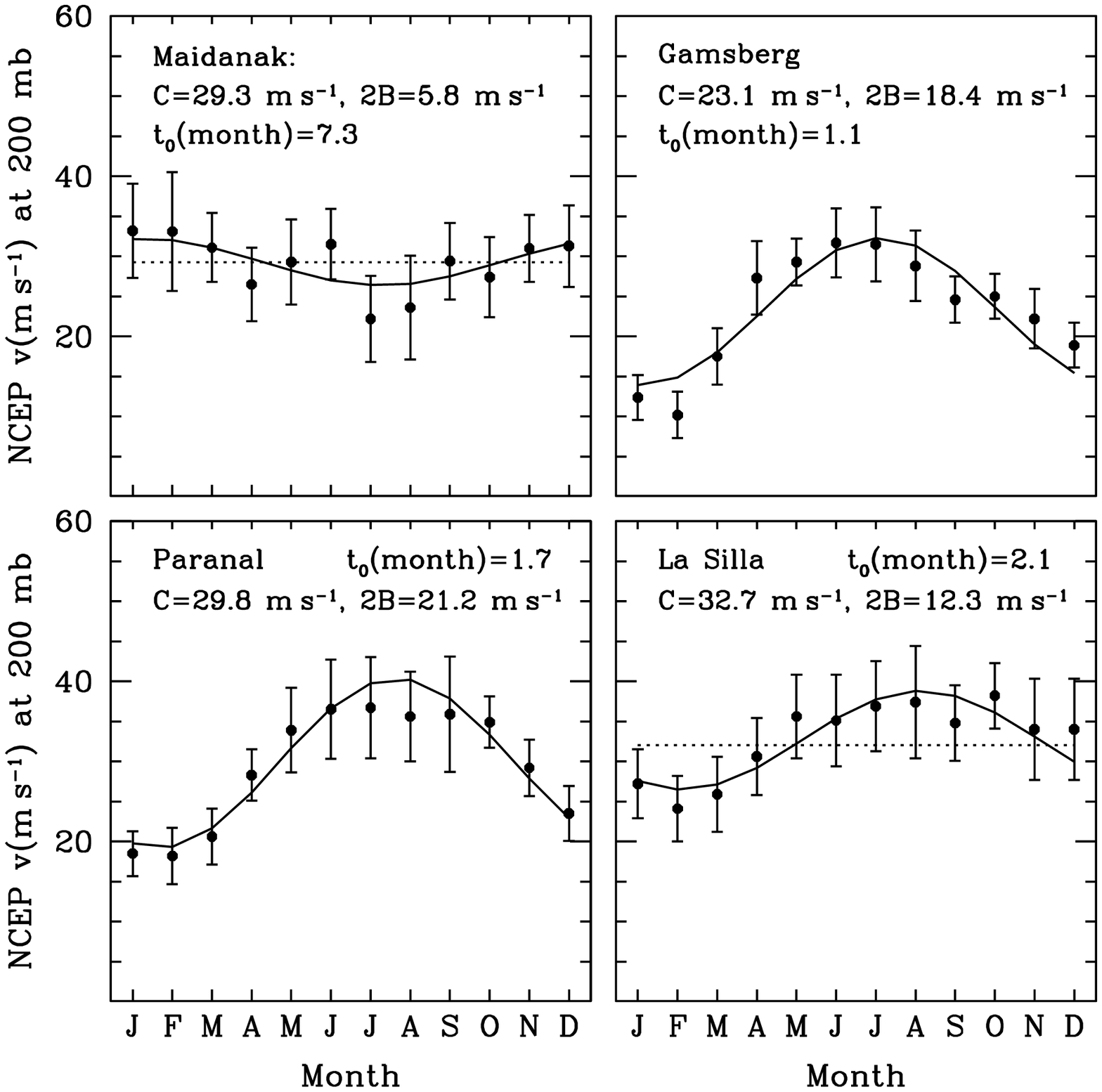}
\caption{Monthly wind speed for the sites considered. Averages and rms 
deviations are indicated by the points and corresponding error bars. The solid 
lines indicate the best fit of a yearly modulated wind speed, 
$v(t) = C_{s} - B\cos\{2\pi (t-t_{0})/1~{\rm year}\}.$ Dotted lines indicate a 
constant wind speed fit, $v(t)=C_{c}$, shown only for the data which do not reject
such hypothesis. \label{ncep_fit2}}
\end{figure}

 Figures \ref{ncep_fit1} and \ref{ncep_fit2} show  plots of the monthly average wind speed at 200$\,$mb 
 for the site indicated, the error bars are the rms fluctuations. The plots are ordered in increasing 
 absolute value of the latitude and grouped depending on the hemisphere: Costa Rica, Sierra Negra, Mauna Kea,
 SPM and Maidanak in the northern hemisphere, and Gamsberg, Paranal and La
 Silla in the southern hemisphere. Costa Rica, where there is not jet stream is shown  as a reference. For 
 the northern sites as the latitute increases so does  the average wind speed,  therefore Sierra Negra 
 presents the lowest average wind at 200~mb of all the sites analyzed, next is Mauna Kea, SPM and Maidanak. 
 Analogously, in the southern hemisphere as the absolute value of the latitude increases so does the average
 wind speed. From the southern sites Gamsberg has the  lowest annual average wind speed at 200~mb. The seasonal
 trend, slowest wind in the summer, is the same for the northern sites Sierra Negra, Mauna Kea and SPM and to
 less extent for Maidanak. Similarly, for the 
sites in the southern hemisphere the months with 
 lower wind  velocity at 200~mb correspond to the southern summer.

    Table \ref{summary} shows a summary of the  wind  velocity annual average. Here we include  
 the uncertainty in the annual average determination, given by the  annual rms fluctuations divided
 by $\sqrt N$ where N is equal to 12. 
Sierra Negra is 3.5$\sigma$, 5.7$\sigma$, 1.6$\sigma$,
 6.9$\sigma$, 7.5$\sigma$  below Mauna Kea, Maidanak, Gamsberg, Paranal and La Silla. The  wind 
 velocity annual average for SPM is  1.2$\sigma$, 1.6$\sigma$, 2.6$\sigma$ below
Maidanak,  
 Paranal and  La Silla respectively. On the other hand, SPM is 1$\sigma$, 1.2$\sigma$, above 
 Mauna Kea and Gamsberg respectively.
For Mauna  Kea, the annual average wind velocity is
 2.4$\sigma$, 2.6$\sigma$, 4.1$\sigma$ below
Maidanak,  Paranal and  La Silla respectively. 
 In contrast, Mauna Kea is 1.6$\sigma$
above Gamsberg,  giving a  statistically tangible  
 site ranking.

     Although annual averages have been computed for all sites it can be shown
that the wind speed is not a constant function of time. Considering that error weighted
averages minimize $\chi^{2}$, we tested the hypothesis of a constant wind speed $v(t)=C_{c}$.
The values obtained are shown in Table~\ref{constante} where  $C_{c}$ are the error  weigthed average
wind speed. The  P$_{11}$ ``hypothesis rejection'' probabilities   are indicated 
for each site. The hypothesis of a constant wind speed is rejected at the 99.9\% confidence level 
for Sierra Negra, San Pedro M\'artir, Mauna Kea, Paranal and Gamsberg.

We also tested the alternative hypothesis of a wind speed following 
\begin{equation}
\label{fit}
v(t) = C_{s} - B\cos\{2\pi (t-t_{0})/1~{\rm year}\}\, 
\end{equation}

with three fitting parameters: $C_{s}$, the average wind speed; $B$ the amplitude
of the wind speed modulation and $t_{0}$ the time of the year of minimum wind speed.
The values of these three parameters, determined through simultaneous $\chi^{2}$ minimization, 
are shown in Table~\ref{modulada}. The corresponding $\chi^{2}$ probabilities are all compatible 
with the hypothesis, which in most cases provides a very good fit to the data. Minimum wind 
speed is reached between June and August in the Northern sites and six months earlier for the 
three Southern sites (Paranal, La Silla and Gamsberg). Even for the sites where a constant
wind speed cannot be rejected (Costa Rica, La Silla and Maidanak), a yearly
modulation provides fits with lower $\chi^{2}$ probabilities. Average 200~mb
wind speeds are only a broad indication of the site properties; the yearly
modulation should be taken into account, as they can be as large as the
average wind speed. We note that $2B\simeq C_{s}$ for Sierra Negra and San Pedro 
M\'artir. The  fits are shown in Figure~\ref{ncep_fit1} and Figure~\ref{ncep_fit2}.


\section{Comparison with Generalized Scidar measurements at SPM}
\label{sec:gs}

In May 2000, measurements of the turbulence profiles $C_\mathrm{N}^2(h)$ and
the  velocity of the turbulent layers $\mathbf{V}_\mathrm{t}(h)$ were
performed during 16 nights at SPM, using the Generalized Scidar (GS) of 
the Laboratoire Universitaire d'Astrophysique de Nice, France. The description 
of the measurements and the results can be found in Avila et al. (2003), 
Avila et al. (2004a) and Avila et al. (2004b). 

   The interest here is two-fold: first, the comparison of the values of 
$v_{200 mb}$ at SPM  for May, obtained from the NCEP/NCAR Reanalysis data base,  with 
the  values of the velocities measured with the GS between
10 and 15 km above sea level, $v_{10;15}$. Second, the comparison of the 
values of  $\mathrm{V}_0$ obtained from the empirical relation proposed by 
Sarazin \& Tokovinin (2002) with the actual values calculated from the  
$C_\mathrm{N}^2$ and  $\mathbf{V}_\mathrm{t}$ profiles measured at SPM,
 $\mathrm{V}_{0;\mathrm{GS}}$.
It is worth mentioning that the GS profiles used in this calculation
do not include the turbulence arising inside the telescope dome.  

As shown in Table 2, for May, the mean and rms values of 
$v_{200 mb}$ are 28.7 and 7.1 m$\,s^{-1}$. The corresponding values of
 $v_{10;15}$ are 26.7 and 12.6  m$\,s^{-1}$. The number of  $v_{10-15}$ values
used for these estimations is 4191. The agreement of both estimates is 
remarkable.

For each simultaneous measurement of $C_\mathrm{N}^2(h)$ and
 $\mathbf{V}_\mathrm{t}(h)$, one value of $\mathrm{V}_{0;\mathrm{GS}}$ is calculated
 by:
\begin{equation}
\label{eq:v0gs}
\mathrm{V}_{0;\mathrm{GS}}=\left[ \frac{\int \mathrm{d}h\;|\mathbf{V}_\mathrm{t}(h)|^{5/3}\; C_\mathrm{N}^2(h)}{\int \mathrm{d}h\;C_\mathrm{N}^2(h)}   \right]^{3/5}.
\end{equation}
This computation was performed for the 3016 measured profiles. The mean and rms values
of $\mathrm{V}_{0;\mathrm{GS}}$ are 12.2 and 5.7  m$\,s^{-1}$, which perfectly agrees 
with  V$_{0}$ = $0.4 v_{200 mb}= 11.5$ m$\,s^{-1}$. This simple comparison supports
the empirical relation proposed  by Sarazin \& Tokovinin (2002).


\section{Conclusions}

      We have analyzed  the seasonal variations of the  monthly average wind velocity
  at 200 mb for
the period between  1980-1995, for Sierra Negra, SPM, Mauna Kea, Paranal, La Silla, 
  Gamsberg, Maidanak and Costa Rica; the latter to be used as a reference. We used the monthly average 
  data based on a 4 records per day,  provided by the NOAA NCEP/NCAR  reanalysis data set.   
  We  have
shown  that Sierra Negra has the lowest yearly average wind velocity of the sites
  analysed. The wind speed at 200$\,$mb in SPM is  comparable to that of Mauna Kea, 
  and in some months probably lower. SPM annual average velocity  is lower  than for Paranal, 
  La Silla  and Maidanak.   Following the criteria that the average  velocity of the 
  turbulence  V$_{0}$ is linearly related to the wind velocity at 200$\,$mb pressure level, que conclude that
  Sierra Negra  and SPM are amongst the best observatory  sites  suitable for
Adaptive  Optics techniques. 
  Furthermore, we have also shown that for all the sites analyzed there is a annual wind speed modulation 
  that should be taken into account when  comparing the quality of different sites.   In particular, the 
  amplitude of the wind speed modulation is comparable to the mean wind speed for Sierra Negra and SPM. 
  
     In the case of SPM, by analyzing V$_{0}$  and $v_{10;15}$ data obtained with a GS during a 16 days 
  campaign, we found that the wind velocity $v_{10;15}$  measured is consistent with the 
  $v_{200 mb}$ value from the NCEP/NCAR reanalysis data base. We obtained for SPM the same  
  relation V$_{0}$ = $0.4 v_{200 mb}$ proposed  by Sarazin \& Tokovinin (2002) for Paranal and 
  Cerro Pach\'on.

\acknowledgements

{NCEP/NCAR Reanalysis data provided by  the NOAA-CIRES Climate Diagnostics Center, Boulder, Colorado,
USA, from their Web site at http://www.cdc.noaa.gov/. R. Avila acknowledges the support of DGAPA, UNAM 
through the grant number IN111403.}

\begin{deluxetable}{c r r r r r}
\tablecaption{Sites Geographic Coordinates \label{sites}
}
\tablewidth{0pt}
\tablehead{
\colhead{Site} & \colhead{Lat} & \colhead{Long} & \colhead{Lat} & \colhead{Long}&\colhead{Elev}\\
\colhead{} & \colhead{} & \colhead{  } & \multicolumn{2}{c}{closest} &\colhead{(m)}}
\startdata
             Costa Rica   &  $+10.00$ & $-85.0$   & $+10.00$ & $-85.00$  & 1100\\

            Sierra Negra &  $+18.98$ & $-97.16$  & $+20.00$ & $+97.50$  & 4580\\
             SPM          &  $+31.04$ & $-115.46$ & $+30.00$ & $-115.00$ & 2890\\
             Mauna Kea    &  $+19.83$ & $-155.47$ & $+20.00$ & $-155.00$ & 4205\\
             Paranal      &  $-24.63$
& $-70.40$  & $-25.00$
& $-70.00$  & 2635\\
             La Silla     &  $-29.25$ & $-70.73$  & $-30.00$ & $-70.00$  & 2400\\
             Gamsberg     &  $-23.34$ & $+16.23$  & $-22.50$ & $+15.00$  & 2347\\
             Maidanak     &  $+38.68$ & $+66.90$  & $+40.00$ & $+65.00$  & 2000 \\
\tableline
\enddata
\end{deluxetable}

\begin{deluxetable}{crrrrrrrrrr}
\tablecaption{Monthly average wind velocity (m$\,s^{-1}$) at 200$\,$mb level for the period 
1980-1995  obtained from the NCEP/NCAR Reanalysis data base
\label{ncep_1}}
\tablewidth{0pt}
\tablehead{
\colhead{M}  & \multicolumn{2}{l}{Costa Rica} & \multicolumn{2}{l}{Sierra Negra} &
\multicolumn{2}{l}{San Pedro M}& \multicolumn{2}{l}{Mauna Kea}& \multicolumn{2}{l}{Paranal}\\
\colhead{} & \colhead{Av} & \colhead{rms} & \colhead{Av} & \colhead{rms} &
\colhead{Av} & \colhead{rms} & \colhead{Av} & \colhead{rms} & 
\colhead{Av}& \colhead{rms} }
\startdata
1 & 13.9 &  3.1 & 26.1  &  5.2 &  32.2   &  4.8   & 29.9 &  4.9 &   18.5 &  2.8    \\           
2 & 13.2 &  2.7 & 27.0  &  6.3 &  35.8   &  7.1   & 32.9 &  4.4 &   18.2 &  3.5  \\  
3 & 12.0 &  3.5 & 27.6  &  7.1 &  38.4   &  9.0   & 33.5 &  5.4 &   20.6 &  3.5   \\ 
4 &  9.4 &  2.9 & 27.8  &  5.4 &  30.2   &  8.3   & 32.0 &  5.9 &   28.3 &  3.2   \\ 
5 &  7.1 &  1.3 & 22.5  &  5.0 &  28.7   &  7.1   & 25.0 &  5.7 &   33.9 &  5.3   \\  
6 &  8.7 &  2.1 & 10.2  &  2.0 &  20.9   &  4.6   & 20.7 &  5.3 &   36.5 &  6.2  \\ 
7 &  8.4 &  2.0 &  7.9  &  1.1 &  10.5   &  3.2   & 18.2 &  3.3 &   36.7 &  6.3   \\ 
8 &  9.9 &  2.3 &  6.9  &  0.9 &  11.9   &  2.5   & 15.8 &  2.2 &   35.6 &  5.6   \\  
9 &  8.9 &  1.7 &  8.1  &  1.6 &  19.7   &  4.6   & 18.1 &  3.3 &   35.9 &  7.2   \\  
10&  9.1 &  1.8 &  12.1 &  2.5 &  26.8   &  4.4   & 19.5 &  3.3 &   34.9 &  3.2  \\  
11&  8.4 &  2.4 & 17.4  &  4.9 &  30.4   &  5.4   & 20.6 &  4.6 &   29.2 &  3.5    \\ 
12& 12.0 &  2.7
& 21.0  &  3.9 &  32.1   &  6.3   & 25.4 &  4.8 &   23.5 &  3.4   \\ 
\tableline
Ave & {\bf 10.1}&2.4& {\bf 17.8} &  4.3 &{\bf  26.5}  &  5.9 & {\bf 24.3} &  4.5 & {\bf 29.3} &  4.7 \\ 
\tableline
\enddata
\end{deluxetable}

\begin{deluxetable}{lrrrrrrrrrrrrrrr}
\tablecaption{Monthly average wind velocity ( m$\,s^{-1}$) at 200$\,$mb level for the period 
1980-1995  obtained from the NCEP/NCAR Reanalysis data base 
\label{ncep_2}}
\tablewidth{0pt}
\tablehead{
\colhead{}  & \multicolumn{2}{l}{La Silla} &  
\multicolumn{2}{l}{Gamsberg}& \multicolumn{2}{l}{Maidanak}\\
\colhead{M} & \colhead{Ave} & \colhead{rms} &
\colhead{Ave} & \colhead{rms} & \colhead{Ave} & \colhead{rms}  
}\startdata
1 &  27.2  &  4.3 &         12.4  &  2.8 &  33.2  &  5.9  \\           
2 &  24.1  &  4.1 &         10.2  &  2.9 &  33.1  &  7.4 \\  
3 &  25.9  &  4.7 &         17.5  &  3.5 &  31.1  &  4.3 \\ 
4 &  30.6  &  4.8 &         27.3  &  4.6 &  26.5  &  4.6 \\ 
5 &  35.6  &  5.2 &         29.3  &  2.9 &  29.3  &  5.3 \\  
6 &  35.1  &  5.7 &         31.7  &  4.3 &  31.5  &  4.4\\ 
7 &  36.9  &  5.6 &         31.5  &  4.6 &  22.2  &  5.4 \\ 
8 &  37.4  &  7.0 &         28.8  &  4.4 &  23.6  &  6.5  \\  
9 &  34.8  &  4.7 &         24.6  &  2.9 &  29.4  &  4.8 \\  
10&  38.2  &  4.1 &         25.0  &  2.8 &  27.4  &  5.0 \\  
11&  34.0  &  6.3 &         22.2  &  3.7 &  31.0  &  4.2  \\ 
12&  34.0  &  6.3 &         18.9  &  2.8 &  31.3  &  5.1 \\ 
\tableline
Ave & {\bf 32.4} &  5.2   & {\bf 23.3}  &  3.6 &  {\bf 29.1} &  5.3\\ 
\tableline
\enddata
\end{deluxetable}

\begin{deluxetable}{c r}
\tablecaption{ Wind speed at 200~mb annual average (1980-1995)
\label{summary}}
\tablewidth{0pt}
\tablehead{
\colhead{Site} & \colhead{ velocity (m$\,s^{-1}$)}}
\startdata
             Costa Rica   &  {\bf 10.1}   $\pm 0.7$\\

            Sierra Negra &  {\bf 17.9}   $\pm 1.25$\\
              SPM         &  {\bf 26.5}   $\pm 1.7$\\
              Mauna Kea   &  {\bf 24.3}   $\pm 1.3$\\
              Paranal     &  {\bf 29.3}   $\pm 1.4$\\
             La Silla     &  {\bf 32.4}   $\pm 1.5$\\
             Gamsberg     &  {\bf 23.3}   $\pm 1.0$\\
             Maidanak     &  {\bf 29.1}   $\pm 1.5$\\
\tableline
\enddata
\end{deluxetable}

\begin{deluxetable}{lcccc}
\tablecaption{Error weighted average wind speeds and $\chi^{2}$ values for the
sites under consideration. $P_{11}$ refers to the $\chi^{2}$ with 11 degrees 
of freedom (12 points - 1 fit parameter) cumulative probability function.
An asterix means rejection of a constant wind speed hypothesis above a 99\%
confidence level.
\label{constante}}
\tablewidth{0pt}
\tablehead{
\colhead{Site} & \colhead{ $C_{c}$ } & \colhead{$\chi^2$} & \colhead{$P_{11}(\leq \chi^{2})$} & \colhead{}\\
\colhead{}     & \colhead {$({\rm m}\,{\rm s}^{-1})$} & \colhead{} & \colhead{}  & \colhead{} }
\startdata
 Costa Rica         & ~~9.19 & ~~9.28 & 0.4037 &   \\
Sierra Negra       & ~~9.23 &  66.15 & 1.0000 & * \\
San Pedro M\'artir &  20.36 &  47.67 & 1.0000 & * \\
Mauna Kea          &  21.35 &  27.98 & 0.9967 & * \\
Paranal            &  26.54 &  36.73 & 0.9999 & * \\
La Silla           &  32.05 &  11.71 & 0.6139 &   \\
Gamsberg           &  21.78 &  52.50 & 1.0000 & * \\
Maidanak           &  29.26 & ~~4.46 & 0.0454 &   \\
\tableline
\enddata
\end{deluxetable}

\begin{deluxetable}{lccccc}
\tablecaption{Error weighted average wind speeds and $\chi^{2}$ values for the
sites under consideration. $P_{9}$ refers to the $\chi^{2}$ with 9 degrees 
of freedom (12 points - 3 fit parameter) cumulative probability function.
All data conform with the hypothesis of a yearly modulated wind speed.
\label{modulada}}
\tablewidth{0pt}
\tablehead{
\colhead{Site}& \colhead{$C_{s}$}& \colhead{2$B$} & \colhead{$t_{0}$} & \colhead{$\chi^2$} & \colhead{$P_{9}(\leq \chi^{2})$}\\
\colhead{}  & \colhead{$({\rm m}\,{\rm s}^{-1})$} & \colhead {$({\rm m}\,{\rm s}^{-1})$} & \colhead{(month)} & \colhead{} & \colhead{}} 
\startdata
Costa Rica         & ~~9.67 & ~~4.21 & 6.82 & 4.55 & 0.1285 \\
Sierra Negra       &  17.37 &  21.47 & 8.00 & 3.12 & 0.0405 \\
San Pedro M\'artir &  25.36 &  24.46 & 7.51 & 6.13 & 0.2729 \\
Mauna Kea          &  24.34 &  16.56 & 8.43 & 0.90 & 0.0004 \\
Paranal            &  29.75 &  21.23 & 1.66 & 2.42 & 0.0169 \\
La Silla           &  32.65 &  12.32 & 2.14 & 2.21 & 0.0122 \\
Gamsberg           &  23.09 &  18.39 & 1.12 & 8.94 & 0.5574 \\
Maidanak           &  29.30 & ~~5.83 & 7.29 & 2.71 & 0.0255 \\
\tableline
\enddata
\end{deluxetable}

\begin{thebibliography}{}

\bibitem[Avila~et~al 2003]{Avila03}
Avila, R., Iba\~nez, F., Vernin, J., Masciadri, E., S\'anchez, L.~J., Azouit, M,
Agabi, A.,Cuevas, S., \&  Garfias, F., 2003,  RevMexAA CS 19, 
San Pedro M\'artir: astronomical site evaluation, 11

\bibitem[Avila~et~al 2004a]{Avila04a}
Avila, R., Masciadri, E., Vernin, J., \&  S\'anchez, L.~J., 2004a,
\pasp, 116, 682
 
\bibitem[Avila~et~al 2004b]{Avila04b}
Avila, R., Vernin, J., Prieur, J.~L., \& Carrasco, E., 2004b, \pasp,
{\it in preparation}

\bibitem[Aviles 2004]{Av01}
Avil\'es, J.L. 2004, Caracterizaci\'on de Sierra Negra para observaciones en el visible.
INAOE M Sc. thesis

\bibitem[Carrasco~et~al 2003a]{carrasco03a} 
Carrasco, E., Carrami\~nana, A., Avil\'es, J.L. \& Yam, O. 2003, \pasp\ 115,879

\bibitem[Carrasco~et~al 2003b]{carrasco03b} 
Carrasco, E., Carrami\~nana, A., Avil\'es, J.L. Yam, O. \& Luna, F. 2003, 
INAOE Technical RT0 548

\bibitem[Carrasco \& Sarazin 2003]{carrasco03c}
Carrasco, E \& Sarazin, M. 2003, RevMexAA CS 19, 103

\bibitem[Chueca et al. 2004]{chueca04}
Chueca, S., Garc\'{\i}a-Lorenzo, B., Mu\~noz-Tu\~non, C. \& Fuensalida, J.J.
Mon. Not. R. Atron. Soc. 2004, 349, 627-631

\bibitem[Cruz-Gonz\'alez~et~al 2003]{cra03}
Cruz-Gonz\'alez, I., Avila, R. \& Tapia M., eds. 2003, RevMexAA CS 19, 
San Pedro M\'artir: astronomical site evaluation 

\bibitem[Kistler~et~al (2001)]{ki01}
Kistler, R., Kalnay, E., Collins, W., Saha, S., White, G., Woollen, J., Chelliah, M., 
Ebisuki, W., Kanamitsu, M., Kousky, V., Van den Dool, H., Jenne, R. \&  Fiorino, M. 2001,
Bull Amer Meteor Soc, 82, 2, 268

\bibitem[Meza~et~al 2004]{Me01}
Meza, J., Torres, A., Aguilar, L. \& Guti\'errez, C. GTM/LMT Technical Report

\bibitem[Michel et al. 2003]{mich2003}
Michel, R., Echevarr\'{\i}a, J., Costero, R., Harris, O., Magall\'on, J. \& Escalante. K. 2003
RevMexAA  32, 2, 291

\bibitem[Michel et al 2003]{mich2003b}
Michel, R., Hiriart, D.,  \& Chapela A., 2003
RevMexAA CS 19, 
San Pedro M\'artir: astronomical site evaluation, 99


\bibitem[Parrao \& Schuster 2003]{parrao03}
Parrao L. \& Schuster, W.J., 2003
RevMexAA CS 19, 
San Pedro M\'artir: astronomical site evaluation, 81

\bibitem[Sarazin \& Tokovinin(2002)]{zt97} Sarazin, M. \& Tokovinin A., 2002,
in Beyond Conventional Adaptive Optics, Venice 7-10, 
eds. R. Raggazoni, N. Hubin, \&  S. Esposito. 
Garching, Germany: European Southern Observatory, 
ESO Conference and Workshop Proceedings, Vol. 58, 321


\bibitem [Schuster, Parrao \& Guichard 2002]{schuster02} 
Schuster, W.J., Parrao, L. \& Guichard, J. 2002, The Journal of Astronomical Data, 8

\bibitem[Roddier 1982]{rod01}
Roddier F.,  Gilli, J.M. \& Lund, G., 1982, J.Op. (Paris), 13, 263

\bibitem[Tapia 2003]{tap03}
Tapia, M. 2003 
RevMexAA CS 19, San Pedro M\'artir: astronomical site evaluation, 75

\end{thebibliography}
\end{document}